\documentclass[12pt]{article}
\usepackage{amsmath,amssymb}
\usepackage{graphicx,color}
\numberwithin{equation}{section}
\usepackage{cite}
\usepackage{bm}
\usepackage{dcolumn}
\newcommand{\be}{\begin{equation}}
\newcommand{\bea}{\begin{eqnarray}}
\newcommand{\eea}{\end{eqnarray}}
\newcommand{\ba}{\begin{array}}
\newcommand{\ea}{\end{array}}
\newcommand{\ee}{\end{equation}}

\expandafter\ifx\csname mathbbm\endcsname\relax

\else

\fi \textheight 22cm \textwidth 15cm \topmargin 1mm \oddsidemargin
5mm \evensidemargin 5mm

\begin{document}
\begin{titlepage}
\hfill \vbox{
    \halign{#\hfil         \cr
         \cr
                      } 
      }  
\vspace*{20mm}
\begin{center}
{\Large {\bf Gauge Symmetry Breaking in Gravity and Auxiliary Effective Action}\\
}

\vspace*{15mm} \vspace*{1mm} {Amin Akhavan}

\vspace*{.4cm}

{\it  School of Particles and Accelerators, Institute for Research in Fundamental Sciences (IPM)\\
P.O. Box 19395-5531, Tehran, Iran\\$email:amin_- akhavan@ipm.ir$ }

\vspace*{2cm}

\end{center}

\begin{abstract}
At first, we consider the path integral method for the covariant symmetry breaking in gravity. We replace the scalar fields, instead of the degrees of freedom which have been removed by gauge fixing constraints. Finally the specific ghost degree of freedom, remains excluded via one of the constraints. Secondly, we define an auxiliary effective action. Entering an auxiliary field, we will have a new dynamic field separated from the fundamental field.

\end{abstract}

\vspace{2cm}

\end{titlepage}

\section{Introduction.a}
Chamseddin and Mukhanov have defined a new kind of the covariant higgs mechanism method in gravity\cite{chams}. By use of the original metric and scalar fields, they defined a new metric as the following,
\be
H^{AB}(x)=g^{\mu\nu}(x)\partial_{\mu}\phi^A(x)\partial_{\nu}\phi^B(x)\nonumber
\ee 
They added a new term functional of $H^{AB}$ to the Einstein and
Hilbert action such that creates the Fierz and Pauli mass terms\cite{fierz}.
The action for the scalar fields that they considered is: 
\be
S(\phi)\propto\int(d^4x\sqrt{-g})\,\,\,3(\frac{1}{16}H^2-1)^2
-\overline{H}^A_{B}\overline{H}^B_{A}\nonumber 
\ee
Such that,$H=H^A_A$ and $\overline{H}^A_B=H^A_B-\frac{\delta^A_B}{4}H$. 

In fact, their symmetry breaking produces a linearly ghost free gravity with one metric in four dimensions. Deser and Boulware had explained a problem in massive gravity\cite{deser}. There is a ghost degree of freedom is coupled to the five degrees of freedom of the massive gravity. To solve the problem, Isham, Salam and Strathdee defined a bigravity with a massless gravity coupled to a massive gravity \cite{isham}, \cite{isham1}. Dvali, Gabadadze, and Porrati presented a gravity on five dimensions with an extra dimension\cite{dvali}. Their gravity on a determined background is free of ghosts. In the letter of Chamseddin and Mukhanov\cite{chams}, A ghost free massive gravity with only one metric without extra dimension, via the symmetry breaking method has been done. And the absorption of the higgs scalars to the massless graviton to generate a massive graviton, has been shown. 

Using the scalar fields for symmetry breaking in gravity field theory, has been explained by t'Hooft\cite{tooft}, at first time. Non zero values of the vacuum expectation of the scalar fields, produced the terms added to the massless gravity action. The produced terms did not have the Fierz-Pauli term, and the ghost degree of freedom could not be excluded. 

Here, we want to introduce the Chamseddin-Mukhanuv mechanism in path integral formulation. The scalar fields is added to the degrees of freedom of the gauge fields. Through considering the expectation values of the scalar fields as a fixed gauge, we present a gauge symmetry breaking field. In section 3, we explain the defined method to obtain a massive vector field. In section 4, we obtain a massive gravity, while considering one of the gauge fixing constraints to eliminate the ghost degree of freedom.       

\section{Introduction.b}
To describe some of the properties of the universe, the scalar fields are very useful. For example C. Armendariz-Picon, V. Mukhanov and
Paul J. Steinhardt have defined the K-essence scalar field\cite{Steinhardt}, an evolving scalar field that produces a negative pressure in the beginning of the matter dominant epoch. In their method, the problem of cosmic coincidence has been resolved without use of fine-tuning. They have used a non-linear kinetic energy density as a functional of $X=g^{\mu\nu}\partial_{\mu}\phi\partial_{\nu}\phi$, to explain the reason of cosmic expansion. If we can find the other fields of the theory (metric field as an example), as the functionals of $X$ or $\phi$, then we will have all the universe based on the only one scalar field and its internal dynamic.

In section 5, we will show the relation between the effective action of the independent fields and the effective action of the dependent fields. Our method is inspired by the paper of R. Fukuda and E. Kyriakopoulos \cite{FUKUDA},they have derived the effective potential through the path integral with a constraint on the zero mode field.

\section{symmetry breaking and the massive vector field}
In the beginning, we want to change the higgs mechanism presentation slightly.
We add a scalar field to a gauge field degrees of freedom in an effective vacuum state.
New degrees of freedom will be presented as a massive vector field. We assume that the
main lagrangian density has been written as a function of a gauge field $A_{\mu}(x)$ and also a complex
scalar field $\phi(x)$:
\be
 \mathcal{L}(A,\phi)=-\frac{1}{2}\partial_{\mu}A_{\nu}\partial^{\mu}A^{\nu}+\frac{1}{2}\partial_{\nu}A_{\mu}\partial^{\mu}A^{\nu}+\frac{m^2}{2}|\partial_{\mu}\phi -iA_{\mu} \phi |^2+...
\ee
If the complex scalar field has been considered like $\phi(x)=\rho(x)e^{i\chi(x)}$,
we will define an effective action in this way:
\be
\Gamma(v,\bar{\rho})=W(J,K)-J^{\mu}v_{\mu}-K\bar{\rho}
\ee
In which:
\be
e^{iW(J,K)}=\int \mathcal{D}A\mathcal{D}\phi\mathcal{D}\phi^{\dag} e^{iS(A,\phi)+i\int d^4xJ^{\mu}(A_{\mu}-\partial_{\mu}\chi)+i\int d^4xK\rho}
\ee
and,
\be
v_{\mu}=\frac{\delta W(J,K)}{\delta J^{\mu}} , \bar{\rho}=\frac{\delta W(J,K)}{\delta K}
\ee
If we consider 
\bea
\langle A_{\mu}(x)-\partial_{\mu}\chi(x)\rangle =0,\nonumber\\
\langle\rho(x)\rangle =1, \nonumber\\
\langle\chi(x)\rangle =\theta(x) 
\eea
as the expectation values in the broken symmetry effective vacuum, 
then the renormalization conditions for a symmetry breaking state, can be defined as:
\bea
\frac{\delta\Gamma(v,\bar{\rho})}{\delta\bar{\rho}}|_{\bar{\rho} =1, v=0}=\frac{\delta\Gamma(v,\bar{\rho})}{\delta v_{\mu}}|_{\bar{\rho} =1, v=0}=0,\nonumber\\
\int dx e^{ik(x-y)}\frac{\delta^2 \Gamma(v,\bar{\rho})}{\delta v_{\mu}\delta v_{\nu}} |_{\bar{\rho}=1,v_{\mu}=0, onshell}=k^2\eta^{\mu\nu}-k^{\mu}k^{\nu}
+m^2
\eea
Next, to obtain the defined effective action, one have to solve the path integral in the equation (2.3). Considering the path integral measure, we will change the field $A_{\mu}(x)$ to the field $V_{\mu}(x)=A_{\mu}(x)-\partial_{\mu}\chi(x)$, and therefore we'll have:
\be
\mathcal{D}A(x)\delta(\partial_{\mu}A^{\mu}(x))\mathcal{D}\phi(x)\mathcal{D}\phi^{\dag}(x)\propto\nonumber\\
\ee
\be
\mathcal{D}V(x)\delta(\partial_{\mu}V^{\mu}(x)-\partial^2\chi(x))\mathcal{D}\chi(x)\rho(x)\mathcal{D}\rho(x)
\ee
Using this transformation, at the first step, we can obtain:
\be 
\langle\partial_{\mu}V^{\mu}(x)-\partial^2\chi(x)\rangle =0
\ee
and also, through the equations (2.5):
\be
\partial^2\theta(x)=0
\ee
In this way, $\theta(x)$ is a fixed classical gauge, in the gauge symmetry breaking mechanism.              

At the second step, via integrating out the field $\chi(x)$, we can remove the Dirac delta function and therefore the path integral measure becomes to this form:
\be
\mathcal{D}V(x)det(\partial^2)\rho(x)\mathcal{D}\rho(x)
\ee
and therefore, one longitudinal degree of freedom from $\chi(x)$ and two vertical degrees of freedom from $A_{\mu}(x)$, have been combined and have formed a spin-one field $V_{\mu}(x)$ with three degrees of freedom.
Finally, by considering this part of the scalar field renormalized lagrangian density:
\be
\frac{m^2}{2}|(\partial_{\mu}-iA_{\mu}(x))e^{i\chi(x)}|^2
\ee
and also through the renormalization condition equations (2.6), the free part of the effective action as a function of $v_{\mu}(x)$ could be presented like this:
\be 
\int d^4x [-\frac{1}{2}\partial_{\mu}v_{\nu}(x)\partial^{\mu}v^{\nu}(x)+\frac{1}{2}\partial_{\nu}v_{\mu}(x)\partial^{\mu}v^{\nu}(x)+\frac{m^2}{2}v_{\mu}(x)v^{\mu}(x)]
\ee
and from here we can obtain a massive spin-one particle in the symmetry breaking renormalized state.

\section{symmetry breaking and the massive gravity}
In this section, as in the previous one, by entering a scalar field into the degrees of freedom of the gravitational metric, the gravitational gauge can be fixed as a selected geometric coordinates. If the symmetric gravitational metric and the scalar field have been presented like $g^{\mu\nu}(x)$ and $\phi^{A}(x)$, we can combine the symmetric metric and the scalar field together to define a new field $G^{AB}(x)$ with additional degrees of freedom. For doing such definition, we use the path integral method like this:
\be
e^{iW(J)}= \int\mathcal{D}G e^{iS(G)+i\int d^4x J_{AB}G^{AB}}\nonumber\\
\ee
\be   
=\int \mathcal{D}g\mathcal{D}\phi\mathcal{D\pi}e^{iS_{g}(g)+i\int d^4x J_{AB}g^{\mu\nu}\partial_{\mu}\phi^{A}\partial_{\nu}\phi^{B}+i\int d^4x\dot{\phi}^A\pi^A-\mathcal{H}_{\phi}(g,\phi,\pi)}
\ee
According to this definition, the relation between expectation values of $g^{\mu\nu}$ and $G^{AB}$ can be written as:
\be
\langle G^{AB}\rangle = \langle g^{\mu\nu}\partial_{\mu}\phi^{A}\partial_{\nu}\phi^{B} \rangle
\ee
Then to define the broken symmetry effective vacuum, the expectation values of fields in such a vacuum,(as like as the equations (2.5),) have to be written like this:
\bea
\langle \phi^A(x)\rangle =X^A(x)\nonumber\\
\langle G^{AB}(x)\rangle =\eta^{AB}
\eea
Such that, $X^A(x)$, is the fixed classical gauge or the fixed geometric coordinates, after symmetry braking mechanism. And for the effective degrees of freedom presentation, we use the fixed coordinates:
\be 
h^{AB}(X^C)=\langle G^{AB}[x^{\mu}(X^C)]\rangle -\eta^{AB}
\ee
For example in four dimensions, $g^{\mu\nu}$ has two degrees of freedom, therefore to obtain a massive metric $G^{AB}$, three degrees of freedom have to be added. In this way, in the definition of a hamiltonian as a function of scalar field, only three degrees of freedom have to be dynamic:
\be 
\mathcal{H}_{\phi}=\sum_{A=1}^3 \dot{\phi}^A\pi^A-\mathcal{L}_{\phi}(\phi,\dot{\phi})
\ee
The effective action, obtains through the partition function -defined in the equation (3.1)- and also renormalization conditions. To obtain an 
effective action in the form of a massive gravity, a part of renormalization condition defined in such a way that the second order terms of 
the effective action could be written like:
\be
\Gamma^{(2)}(h)=\partial_{C}h^{AC}\partial{A}h_{B}^{B}-\partial_{C}h^{AC}\partial_{B}h_{A}^B+\frac{1}{2}\partial_{C}h_{AB}\partial^{C}h^{AB}
-\frac{1}{2}\partial_{C}h_{A}^{A}\partial^{C}h_{B}^{B}-mh_{A}^{A}h_{B}^{B}+mh_{A}^{B}h_{B}^{A}
\ee
In this form of the effective action, because of Fierz-Pauli massive terms, we can find five dynamics equations of motion and five constraint equations of motions. And also we can have five dynamics renormalized degrees of freedom and five auxiliary renormalized degrees of freedom.
But, it should be noted that, with a very little change in the massive terms, six dynamics degrees of freedom and four auxiliary degrees of
freedom will be created. 

To obtain the complete action, as like as the previous section, we have to attend on the measure of the path integrals. In the second path integral
in the equation (3.1), the gauge fixed measure is:
\be
\mathcal{D}g\prod_{A=1}^{3}\mathcal{D}\phi^A\mathcal{D}\pi^A \prod_{A=1}^{4}\delta(\mathcal{F}^A(g,\phi))det(\frac{\delta\mathcal{F}^A}{\delta\epsilon^{\mu}})
\ee 
by defining $\mathcal{F}^A(g,\phi)=0$, as the gauge fixing constraints. Now, with the use of the variable change 
$G^{AB}=g^{\mu\nu}\partial_{\mu}\phi^A\partial_{\nu}\phi^B$,
and also through the integrating out of the scalar fields and the Dirac delta functionals, we can face with the measure of the first path integral in the equation (3.1):
\be  
\mathcal{D}G[det(\frac{\delta G^{AB}}{\delta g^{\mu\nu}})]^{-1}\delta(\mathcal{F}^4(G,\overline{\phi}))det(\frac{\delta\mathcal{F}^A(G,\phi)}{\delta\epsilon^{\mu}})|_{\phi=\overline{\phi}}
\ee
so that,
\be
\mathcal{F}^{1,2,3}(G,\overline{\phi})=\frac{\delta\mathcal{L}_{\phi}}{\delta\phi^4}|_{\phi=\overline{\phi}}=0,\nonumber\\
\ee
\be
[det(\frac{\delta G^{AB}}{\delta g^{\mu\nu}})]^{-1}=\int\mathcal{D}\chi e^{-\chi Q^2\chi},\,\,\,\,\,\,\,\,Q^{AB\,\,\,\,CD}=\partial_{\mu}\overline{\phi}^A\partial_{\nu}\overline{\phi}^B\delta^{\mu C}\delta^{\nu D}
\ee

In this way, a sort of massive quantum gravity has been approached that has five bare degrees of freedom, because of the Dirac delta functional in the equation (3.8). But there is a problem here: a very little displacement in the massive terms in the renormalized condition equation (3.6), increases the number of the renormalized degrees of freedom to six degrees, but such a displacement doesn't have any effect on the number of the bare degrees of freedom. In the other words, the elimination of the sixth renormalization degree of freedom is because of a constraint equation in the effective equations of motion(such a constraint equation goes to dynamic one after a little change in the massive terms.), while the elimination of the sixth bare degree of freedom is because of a constraint equation apart from the equations of motion, $\mathcal{F}^4(G)=0$.To solve the problem that has been described and to define an equal representation between the renormalized degrees of freedom and the bare degrees of freedom, we consider two propositions:

a. Massive terms in the action $S_{G}$ in the equation (3.1), have to be written in the form of Fierz-Pauli terms.

b. We define $\mathcal{F}^4(G)$ such that any metric $G_{AB}(x)$ that satisfies the constraint equations of Fierz-Pauli action, satisfies the equation $\mathcal{F}^4(G)=0$ as the same.

\section{Auxiliary effective action}
In this section, we want to define a new effective field theory which has new effective fields in addition to the fields in the fundamental action.
For this purpose, we will define a new partition function upon a new effective vacuum state. At first, the partition function over the fundamental fields is as shown bellow:
\be 
e^{iW(J,K)}=\int\mathcal{D}\phi e^{iS(\phi,X(\phi))+iJ\phi+iKX(\phi)}
\ee
Therefore we can write the partition function as follow:
\be
e^{iW(J,K)}=\int\mathcal{D}\phi\mathcal{D}X e^{iS(\phi,X)+iJ\phi+iKX}\delta(X-X(\phi))
\ee
If we rewrite the Dirac delta like, $\int\mathcal{D}Z e^{iZ(X-X(\phi))}$, we will have:
\be
e^{iW(J,K)}=\int\mathcal{D}Z e^{iV(J,K,Z)}
\ee
In this way,
\be
e^{iV(J,K,Z)}=\int\mathcal{D}\phi\mathcal{D}X e^{iS(\phi,X)+iJ\phi+iKX+iZ(X-X(\phi))}
\ee
In the above equations, $Z(x)$ is an auxiliary field. The path-integration over such a field, determines the amount of the field $X(x)$, from $X[(\phi(x))]$ as a functional of the field $\phi(x)$. In the other words, considering all of the values of the field $Z(x)$, is equivalent to a complete set of facilities of measurement of the field $X(x)$ as a functional of the field $\phi(x)$.

Now, we define an effective measurement has been done in such a way that, the obtained value of the field $X(x)$ is deviated from the primary functional of the field $\phi(x)$. In fact, the method of the defined measurement doesn't have a complete set of facilities. In such a measurement, the effective values of the field $X(x)$ is independent from the values of the field $\phi(x)$. In this case, we can define a partition function that doesn't have a path integration over The field $Z(x)$, and for this reason, the field $X(x)$ is free from the field $\phi(x)$. Such a partition function as a functional of the determined auxiliary field $Z(x)$, is defined in the equation (4.4).

To choose the determined field $Z(x)$ in this partition function, it should be noted that if the effective measurement of the field $X(x)$ is independent from the field $\phi(x)$, but the expectation values of the field $X(x)$ have to be dependent, in the auxiliary vacuum state:
\be
\langle \Omega_{Z}|X(x)|\Omega_{Z}\rangle = \langle \Omega_{Z}|X[\phi(x)]|\Omega_{Z}\rangle
\ee  
In this case, through the equation (4.4), we will have:
\be 
\frac{\delta V(J,K,Z)}{\delta Z}|_{J=K=0}=\langle \Omega_{Z}|X(x)-X[\phi(x)]|\Omega_{Z}\rangle=0
\ee
Therefore, the determined field $Z(x)$ is the extremum point of the defined partition function.

Also we define a new effective action named auxiliary effective action as follow:
\be
\Gamma(\varphi_{Z},\overline{X}_{Z},Z)=V(J,K,Z)-J\varphi_{Z}-K\overline{X}_{Z}
\ee
while the effective fields have been defined:
\be
\overline{X}_{Z}=\frac{\delta V(J,K,Z)}{\delta K},\,\,\,\,\,\varphi_{Z}=\frac{\delta V(J,K,Z)}{\delta J}
\ee
Thus, the extremum of the partition function is the extremum of the effective action, as the same:
\be
\frac{\delta\Gamma(\varphi_{Z},\overline{X}_{Z},Z)}{\delta Z}=\frac{\delta V}{\delta J}\frac{\delta J}{\delta Z}+\frac{\delta V}{\delta K}\frac{\delta K}{\delta Z}+\frac{\delta V}{\delta Z}-\frac{\delta J}{\delta Z}\varphi_{Z}-\frac{\delta K}{\delta Z}\overline{X}_{Z}\nonumber\\
\ee
\be
=\frac{\delta V(J,K,Z)}{\delta Z}=0
\ee
And then, for computing the effective action, we have to prescribe related renormalization conditions. The renormalization conditions in the first and the second order are:
\be 
\frac{\delta\Gamma(\varphi_{Z},\overline{X}_{Z},Z)}{\delta\varphi_{Z}}|_{\varphi_{Z}=0,\overline{X}_{z}=0}=
\frac{\delta\Gamma(\varphi_{Z},\overline{X}_{Z},Z)}{\delta\overline{X}_{Z}}|_{\varphi_{Z}=0,\overline{X}_{z}=0}=0
\ee
And also,
\be
\int dx e^{i p (x-y)} \frac{\delta ^2\Gamma(\varphi_{Z},\overline{X}_{Z},Z)}{\delta\varphi_{Z}(x)\delta\varphi_{Z}(y)}|_{\varphi_{Z}=0,\overline{X}_{z}=0,p^2 \rightarrow m_{\phi}^2}\rightarrow p^2-m_{\phi}^2\nonumber
\ee
\be
\int dx e^{i p (x-y)}\frac{\delta^2\Gamma(\varphi_{Z},\overline{X}_{Z},Z)}{\delta\overline{X}_{Z}(x)\delta\overline{X}_{Z}(y)}|_{\varphi_{Z}=0,\overline{X}_{z}=0, p^2 \rightarrow m_{X}^2}\rightarrow p^2-m_{X}^2.\ee

For more conditions, we can consider the higher order derivatives of the effective action as the vertices in the Feynman diagrams.

Finally, we can find the relation between the expectation values in the fundamental vacuum state and the expectation values in the auxiliary vacuum state. Using the equations (4.3),(4.8) we will have:
\be  
\langle \Omega|\phi(x)|\Omega\rangle=\frac{\delta W(J,K)}{\delta J}\nonumber\\
\ee
\be
=e^{-iW(J,K)}\int\mathcal{D}Z e^{iV(J,K,Z)}\frac{\delta V(J,K)}{\delta J}=\frac{\int\mathcal{D}Z e^{iV(J,K,Z)}\varphi_{Z}}{\int\mathcal{D}Ze^{iV(J,K,Z)}}
\ee
And also in the same method:
\be
\langle\Omega|X[\phi(x)]|\Omega\rangle=\frac{\int\mathcal{D}Z e^{iV(J,K,Z)}\overline{X}_{Z}}{\int\mathcal{D}Ze^{iV(J,K,Z)}}
\ee
And therefore we can say that in fact, the fundamental expectation value is the average of the values obtained from all effective measurements.

\section{Conclusion}
In this paper, The first purpose was adding the scalar degrees of freedom to the set of a gauge field, per unit quantum state. In fact, by using the path integration over the scalar fields, we have restored the gauge degrees of freedom which had been removed by gauge fixing mechanism. In the massless gravity, one of the removed degrees of freedom is ghost, and we didn't have to restore it. Therefore one of the gauge fixing constraints, remains in the massive gravity partition function. But of course the remained constraint is compatible with the constraint equations of motion of Fierz-Pauli massive gravity. 

The second subject we have studied, was obtaining new measurable fields from the fundamental scalar field of the theory. We defined a new field as a functional of the fundamental field and we considered a situation of measurement, in which, the results of the new field are independent values. Assume that In the fundamental situation of measurement, the observed values are not separated. But the question is: in what case of observation, the universe is formed by some independent fields?


\vspace*{1cm}



\begin{thebibliography}{99}

\bibitem{chams}
A.H.Chamseddine, V.Mukhanov, Higgs for graviton: simple and elegant,
JHEP 1008, 011 (2010).
\bibitem{fierz}
M. Fierz and W. Pauli, Proc. Roy. Soc. Lond. A 173, 211 (1939).
\bibitem{deser}
D. G. Boulware and S. Deser, "Can gravitation have a finite range?," Phys. Rev. D 6, 3368
(1972).
\bibitem{isham}
C. J. Isham, A. Salam and J. A. Strathdee, "F-dominance of gravity," Phys. Rev. D 3, 867
(1971).
\bibitem{isham1}
A. H. Chamseddine, A. Salam and J. A. Strathdee, "Strong gravity and supersymmetry,"
Nucl. Phys. B 136 (1978) 248.
\bibitem{dvali}
G. R. Dvali, G. Gabadadze and M. Porrati, "4D gravity on a brane in 5D Minkowski space,"
Phys. Lett. B 485, 208 (2000).
\bibitem{tooft}
G. 't Hooft, "Unitarity in the Brout-Englert-Higgs mechanism for gravity," arXiv:0708.3184.
\bibitem{englert}
R.Brout, F.Englert, E.Gunzig, The creation of the universe as a quantum phenomenon,
Annals of Physics. Vol 115, page 78.
\bibitem{FUKUDA}
R. Fukuda, E. Kyriakopoulos, Derivation of the effective potential,
Nuclear Physics B85 (1975) 354-364.
\bibitem{percacci}
R.Percacci, The higgs phenomenon in quantum gravity, Nuclear Physics B. Vol
353, page 271.
\bibitem{CONSTRAINT}
L. O'Raifeartaigh, A. Wipf and H. Yoneyama, The constraint effective
potential, Nuclear Physics B271 (0986) 653-680.
\bibitem{Wetterich}
A. Ringwald,  C. Wetterich, Average action for the N-component
$\varphi^4$ theory, Nuclear Physics B334 (1990) 506-526.
\bibitem{Steinhardt}
C. Armendariz-Picon, V. F. Mukhanov, and P. J. Steinhardt, A Dynamical solution to the
problem of a small cosmological constant and late time cosmic acceleration, Phys.Rev.Lett.
85 (2000) 4438-4441.
\bibitem{kurt}
Kurt Hinterbichler, Theoretical aspects of massive gravity, arXiv:1105.3735
[hep-th] 2 Oct 2011.
\bibitem{afshordi}
 M.Saravani,N.Afshordi,R.B.Mann, Dynamical Emergence of Universal Horizons during
the formation of Black Holes, Phys. Rev. D 89, 084029 (2014).




\end{thebibliography}
\end{document}